\documentclass[letterpaper, 10 pt, conference]{ieeeconf}  % Comment this line out
\IEEEoverridecommandlockouts                              % This command is only
\usepackage{graphics} % for pdf, bitmapped graphics files
\usepackage{epsfig} % for postscript graphics files
\usepackage{times} % assumes new font selection scheme installed
\usepackage{amsmath} % assumes amsmath package installed
\usepackage{amssymb}  % assumes amsmath package installed
\usepackage{subcaption}
\usepackage{hyperref}
\usepackage{tabularx}
\usepackage{multirow}

\let\labelindent\relax

\usepackage{enumitem}

\usepackage{color}

\title{\LARGE \bf
EdgeBench: Benchmarking Edge Computing Platforms}

\author{Anirban Das, Stacy Patterson, Mike P. Wittie% <-this % stops a space
 \thanks{Anirban Das and Stacy Patterson are with the Department of Computer Science, Rensselaer Polytechnic Institute, Troy, New York 12180, USA. {\tt\small dasa2@rpi.edu, sep@cs.rpi.edu}}% <-this % stops a space
\thanks{Mike P. Wittie is with the Gianforte School of Computing at Montana State University, Bozeman, Montana 59715, USA. {\tt\small mwittie@cs.montana.edu}}% <-this % stops a space
\thanks{This work was funded in part by NSF awards CNS-1527287, CNS-1553340, CNS-1555591, and CNS-1527097}
}

\begin{document}

\maketitle
\thispagestyle{empty}
\pagestyle{empty}

%%%%%%%%%%%%%%%%%%%%%%%%%%%%%%%%%%%%%%%%%%%%%%%%%%%%%%%%%%%%%%%%%%%%%%%%%%%%%%%%
\begin{abstract}
The emerging trend of edge computing has led several cloud providers to release their own platforms for performing computation at the `edge' of the network. 
We compare two such platforms, Amazon AWS Greengrass and Microsoft Azure IoT Edge, using a new benchmark comprising a suite of performance metrics. We also compare the performance of the edge frameworks to cloud-only implementations available in their respective cloud ecosystems.
Amazon AWS Greengrass and Azure IoT Edge use different underlying technologies, edge Lambda functions vs. containers, and so we also elaborate on platform features available to developers.
Our study shows that both of these edge platforms provide comparable performance, which nevertheless differs in important ways for key types of workloads used in edge applications. 
%Further our study shows edge computing is much cheaper and bandwidth conserving compared to cloud alternatives. 
Finally, we discuss several current issues and challenges we faced in deploying these platforms.

\end{abstract}

%%%%%%%%%%%%%%%%%%%%%%%%%%%%%%%%%%%%%%%%%%%%%%%%%%%%%%%%%%%%%%%%%%%%%%%%%%%%%%%%
\section{INTRODUCTION} 
\label{sec.introduction}

As the Internet of Things~(IoT) is becoming mainstream, the number of connected devices is growing at an exponential rate~\cite{gartner21billion}. 
In this paradigm, IoT devices, which are often geographically distributed at the edge of the network, will generate a massive quantity of data. 
Transmitting, storing, and processing this huge amount of data in the cloud is expected to lead to high bandwidth usage and prohibitive costs~\cite{edgevision}.
Further, many  applications that run at the edge of the network, such as autonomous vehicles and augmented reality, have real-time requirements that are difficult to meet with relatively distant cloud datacenters~\cite{etsiedgecomputing}. 

Edge computing has the potential to mitigate these cost and performance bottlenecks. 
This computing paradigm enables applications to leverage compute nodes in close proximity to data sources to perform data processing, such as  aggregation, filtering, and classification, before forwarding the results to other nodes and cloud servers~\cite{edgevision, satyanarayanan2017emergence}. 
For example, instead of sending an image to the cloud to perform facial recognition, and edge device may simply report whether the image contains the particular signature.
This approach reduces bandwidth usage and can  speed up application response.

To make application development in the edge computing model easier, several cloud providers,
% such as Amazon, Microsoft, IBM and recently Google, 
have put forward their own edge computing platforms. 
These platforms provide the ability to deploy and orchestrate applications, such as machine learning models, on edge devices in the form of stateless serverless functions or user code in containers. 
%User code is deployed as a stateless function, and 
The resource provisioning and runtime is provided by the edge platforms. 
Such serverless functions or user scripts in containers then act on raw data, and depending on the configuration, send results to the cloud and additionally also make them available in other cloud services. 

Since these platforms use different paradigms and technologies, it is important to compare them  to understand their tradeoffs and to select the best platform for a given use case. 
Criteria of interest include the platform architecture, programmability, performance, and cost.
To quantify these differences consistently requires benchmarks of common uses cases.
Further, it requires standardized collection of metrics, such as end-to-end latency, device compute time, device resource utilization, bandwidth usage, and cost.
To make an informed choice between edge and cloud platforms, the benchmarks and metrics must also allow fair comparison across different types of deployments. 

We present EdgeBench\;\footnote{\url{https://github.com/akaanirban/edgebench}} -- an open-source benchmark suite for serverless edge computing platforms.
EdgeBench features three key applications:
a speech/audio-to-text decoder, 
an image recognition machine learning model, 
and a scalar value generator emulating a sensor. 
Each application processes a bank of input data on an edge device and sends results to cloud storage.
We target EdgeBench for two of the most popular edge computing platforms currently available, AWS Greengrass~\cite{aws_dev_guide} and Microsoft Azure IoT Edge~\cite{azure_dev_guide}. 
EdgeBench also provides cloud-based workload implementations.
Our aim is to quantify the differences between the different edge platforms, as well as the providers' respective cloud-only alternatives.
In future work, we plan to extend EdgeBench to other emerging edge platforms, such as Google's Cloud IoT Edge~\cite{googleiotedge} and IBM Watson IoT Platform Edge~\cite{watsoniotedge}, as these offerings mature.
We report on initial experiments with  EdgeBench using  a Raspberry Pi 3B,  a relatively resource-constrained device, to emulate the IoT device that sends traffic to the AWS and Azure cloud platforms. We provide a performance comparison across the different workloads and platforms.

 EdgeBench complements previous work on benchmarking serverless cloud computing platforms.
Malawski et al. have developed two CPU-intensive benchmark suites and evaluated them on the different industry providers~\cite{malawski2017benchmarking}.
The recent work by McGrath and Brenner presents a comparison of the overhead of various platforms, measured using a custom-developed tool~\cite{serverlessicdcs}.
The work by Back and Andrikopoulos presents a performance study of industry serverless cloud platforms using compute/memory constrained workloads, with a  focus on cost~\cite{microbenchmark}.
Finally, Deese presents a study of the performance of $K$-means clustering using AWS Lambda functions in the cloud~\cite{Deese18}.
To the best of our knowledge, our work is the first that benchmarks industry platforms that use  the serverless paradigm for edge computing.

The rest of the paper is organized as follows. 
Sec.~\ref{sec.sysarchitecture} provides details about the architectures of AWS Greengrass and Azure IoT Edge. 
Sec.\ref{sec.benchmarks} describes the EdgeBench workloads and metrics. 
In Sec.~\ref{sec.setup_results}, we describe the experimental setup and results of the benchmark study
%In Sec.~\ref{sec.discussion} we discuss 
and discuss some observations, and finally, we conclude in Sec.~\ref{sec.conclusion}.

%%%%%%%%%%%%%%%%%%%%%%%%%%%%%%%%%%%%%%%%%%%%%%%%%%%%%%%%%%%%%%%%%%%%%%%%%%%%%%%%
\section{SYSTEM ARCHITECTURES} \label{sec.sysarchitecture}
On an abstract level, both platforms, AWS Greengrass (henceforth, Greengrass)  and Azure IoT Edge (henceforth, Azure Edge) share a common general architecture. 
%There is an edge device connected to cloud services. At first, respective platform's edge run-times need to be deployed on this device in order to let the user run Lambda functions or code in Docker containers. 
There is an edge device that runs user code in the platform's runtime system. 
%\sq{Is it environment or system? We should use the long form description, whatever it is, instead of the shorthand `runtime'}. 
This user code can access local volumes or local devices, such as sensors and cameras, performs computations, and sends messages to the cloud. 
In both platforms, the cloud has a high throughput message ingestion service, the IoT Hub.
The cloud ingests the messages from the edge devices and sends them to configurable destinations, such as AWS~S3 or Azure~blob for storage using `Rule' (for AWS) or a `Route' (for Azure). 
% A `route' or a `rule' lets the cloud system redirect the incoming messages to user defined endpoints based on user defined logic.

Below, we highlight some of the details of each platform:
\subsection{AWS Greengrass}
\begin{figure}[t]
      \centering
      \includegraphics[width=0.9\linewidth]{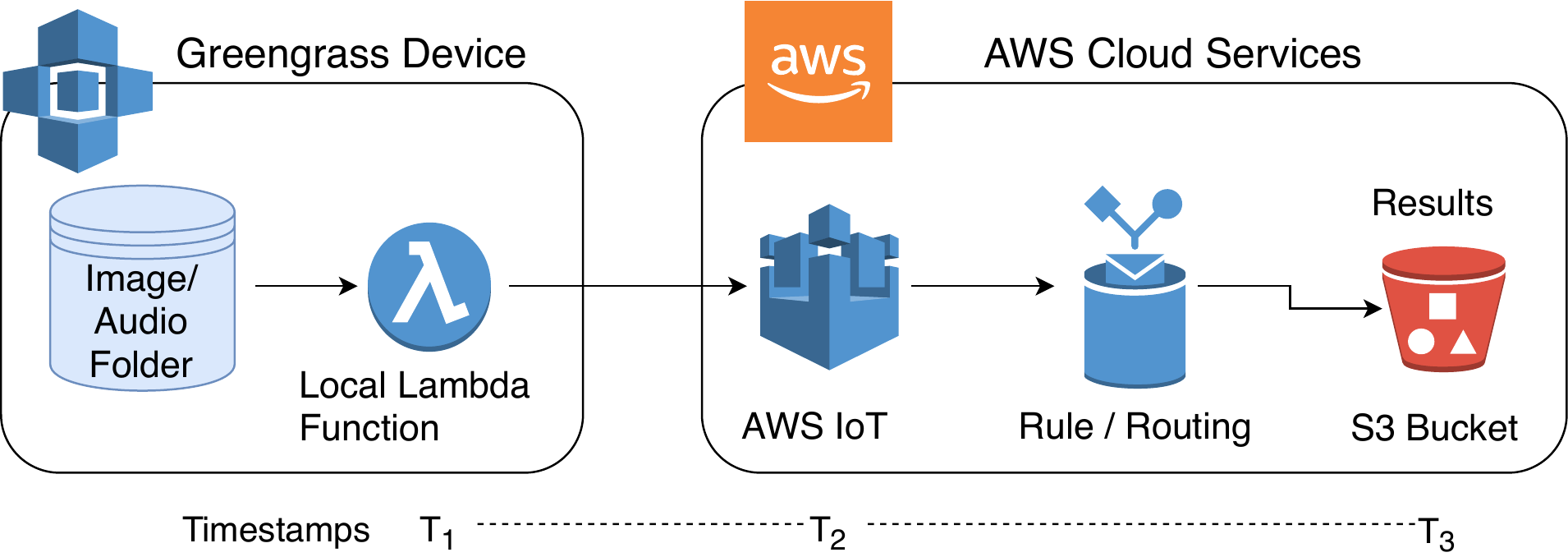}
      \caption{Amazon Greengrass Architecture.}
      \vspace{-10pt}\label{fig.Greengrass_architecture}
\end{figure}
The Greengrass pipeline is shown in Fig.~\ref{fig.Greengrass_architecture}.
Greengrass edge devices run the Greengrass core software. The core software allows users to run Lambda functions locally on the edge devices and manage, modify or update them through the AWS console website or deployment API. Developers can constrain the maximum memory usage of the local Lambda functions.
The Greengrass core software also takes care of the  authentication, authorization, and secure message routing, through the MQTT protocol~\cite{mqtt}, between the devices, Lambda functions, and the cloud.

 Greengrass core Lambda runtime currently supports code deployment in Python 2.7, Node.JS 6.10, Java 8, C, C++, and any language that supports importing C libraries. Code running inside Lambda functions can also access all other AWS services, such as Amazon S3 or DynamoDB, using the standard AWS SDKs. 
Once the AWS IoT Hub receives a message in the cloud,  a `Rule' can be defined to trigger one of 15 actions (as of now), including invoking Lambda functions that run in the cloud or saving data in S3 or DynamoDB. 
If the `Rule' declares to save messages in S3 storage, the hub does so by creating one blob file for each message in the specified S3 bucket, as soon as the message is processed by the hub.

\subsection{Microsoft Azure IoT Edge}
\begin{figure}[t]
      \centering
      \includegraphics[width=0.9\linewidth]{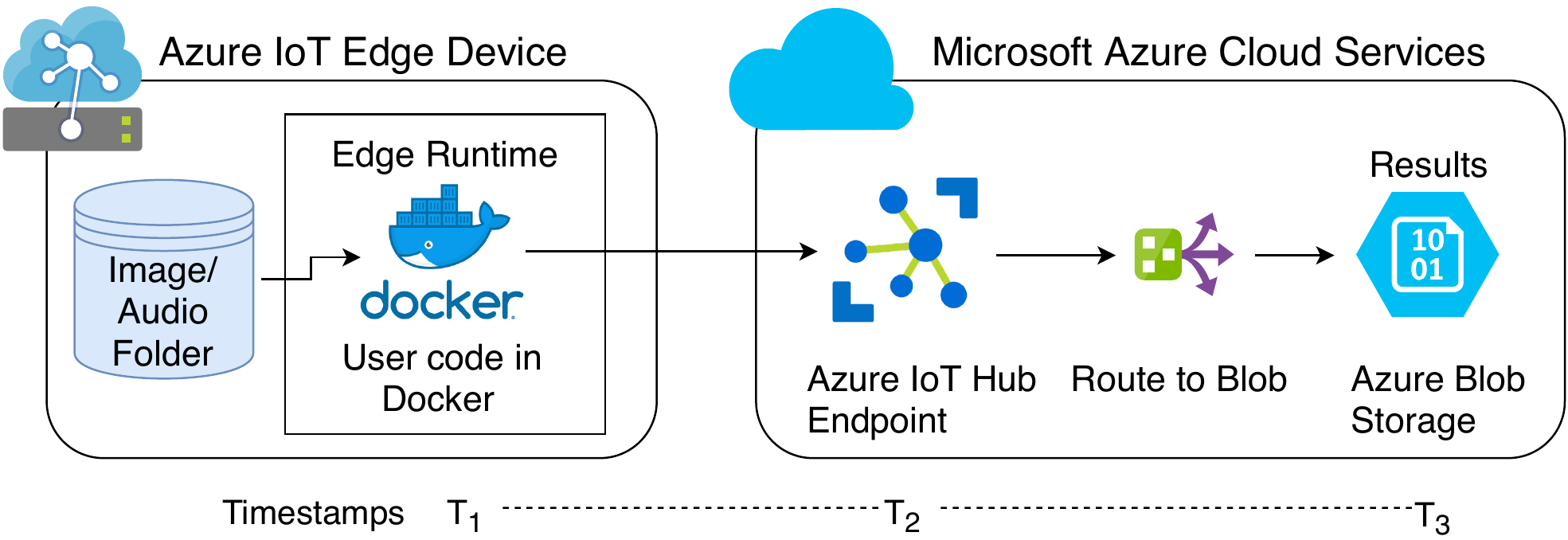}
      \caption{Azure IOT Edge Architecture.}
      \vspace{-10pt}\label{fig.azure_iot_architecture}
\end{figure}

Azure Edge uses lightweight virtualization, specifically, Docker compatible containers, to deploy computation on edge devices. 
The Docker containers run as `edge modules' in the Azure Edge platform, as shown in  Fig.~\ref{fig.azure_iot_architecture}. 
The modules can contain  Azure Functions, user code, and libraries. As of now, the platform  supports five languages: C\#, C, Node.JS ver $>$ 0.4.x.x, Python (both 2.7 and 3.6), and Java 7+.  
It is also possible to deploy Streaming Analytics and Azure~ML models directly in the containers. The former is a managed service from Azure for doing analytics on streaming data and the latter are the models developed in Azure's machine learning service. Modules can be deployed, updated, and modified via the Azure IoT Edge Cloud web interface or the Azure command line interface. 

The runtime system on a single edge device consists of an \texttt{edgeAgent} module and an \texttt{edgeHub} module. The \texttt{edgeAgent} takes care of provisioning and monitoring user deployed modules. The \texttt{edgeHub} takes care of the connection between the modules and the cloud and also maintains security and authentication. The \texttt{edgeHub} supports edge-to-cloud connections using the MQTT and AMQP protocols 
The edgeAgent sends messages to the cloud-hosted Azure IoT Hub, as shown in Fig.~\ref{fig.azure_iot_architecture}.
%sep: I rewrote this in the passive voice and I like it that way :)
Messages are then routed from the IoT Hub to a user-specified IoT Hub Endpoint, such as Azure Blob Storage. 
For the Blob Storage Endpoint, the IoT Hub batches the incoming messages and writes multiple results in a single blob file. If user selects Blob Storage endpoint, batching is the only option. 
The batching window can be configured by either time window, the smallest being 60\;s, or by chunk size, the smallest being 10\;MB.

%%%%%%%%%%%%%%%%%%%%%%%%%%%%%%%%%%%%%%%%%%%%%%%%%%%%%%%%%%%%%%%%%%%%%%%%%%%%%%%%
\section{EDGEBENCH} \label{sec.benchmarks}
In this section, we describe EdgeBench benchmark suite and the performance metrics. We also summarize the cloud-based implementations of the benchmark applications.

\subsection{Benchmark Applications and Workloads} \label{benchmark_apps.sec}

We selected three canonical applications: a speech/audio-to-text  application;  an image recognition application; and a scalar value generator that emulates a sensor, e.g. a temperature sensor. 
With the popularity hike in use of a myriad of smart speakers, such as Amazon Echo and Google Home, has made audio and speech decoding and translation very relevant. Similarly, with the proliferation of smart cameras and autonomous vehicles, image processing and image classification has become very common. Currently, these applications are often executed in the cloud. Hence, it is interesting to investigate performance of such applications in an edge computing setting.
 The scalar benchmark, however, is an example of an extremely lightweight workload; it allows us to measure the performance of each framework when the computation and data volume at the edge are negligible. 
 
All benchmark codes are written in Python. In all three pipelines, the edge devices send data in messages to the IoT Hub
% sep: This is not redundant because there can be a variety of rules or routes, we need to say which one we use.
We use a either a `Rule' or `Route' to push each  message payload in the cloud to an AWS S3 bucket or an Azure Blob, respectively.

\begin{itemize}[topsep=0mm,leftmargin=*]
\item \textbf{Audio/Speech to Text Translation (Audio Pipeline)}
%The final benchmark application consists of an audio/speech to text translation pipeline. 
Here, the edge application reads audio files from a local directory, decodes them to get the translated text, and then sends the text to the cloud. Each audio file is processed one at a time.
 For our experiments, we use a mobile version of Carnegie Mellon University's Sphinx speech recognition system, called PocketShpinx~\cite{pocketsphinx}. We use the default acoustic model provided with the package. For the audio workload, we use 104 samples contributed by user `\textbf{rhys\_mcg}' in Tatoeba Corpus~\cite{tatoeba}, a free collaborative online database of example sentences. 
%Each file in this database is the audio of the speaker uttering a full sentence. 
We have converted the audio files into 16khz, 16 bit, mono `wav' file format  to comply with the requirements of PocketSphinx. 
The average realtime length of files in the dataset is about 2.4\;s.

\item \textbf{Image Recognition (Image Pipeline):}
The serverless function performs an image recognition task; specifically, given an image as input, the function recognize the objects present in the image and generates class labels for these objects. In both platforms, the edge application reads an image from a directory. 
It then uses OpenCV~\cite{opencv} to  resize the image to standard ($224\times224\times3$) size. Finally, the application uses the open-source, deep learning framework MXNet~\cite{MXNet} to recognize and classify the objects in that image. The results are sent to the cloud. 
This is repeated for all images in the directory.
For the classifier architecture, we chose a pretrained classifier,
%(trained on Imagenet ILSVRC2012 dataset \cite{imagenet2012})
Squeezenet~\cite{squeezenet}  because its small model size ($\approx$ 5\;MB) and low compute footprint are suitable for resource-constrained edge devices.  For the input workload, we select 500 images from the ILSVRC2012 image dataset~\cite{imagenet2012}. The input is stored on a local directory on the device.  
%This local directory can be easily replaced by a web-cam or other similar device for more realistic image capture scenarios.

\item \textbf{Scalar Sensor (Scalar Pipeline):}
The application is a simple sensor emulator. 
The serverless function  generates random scalar values at a user-specified frequency. At a user-specified interval, e.g., 1s, the set of values generated in that interval is sent to the cloud
in a single message. The cloud side of the pipeline then simply stores these values in the specified storage location.
%each message is a list of few temperature and humidity values. After deployment, the edge device sends these messages to the cloud for aggregation or storage. The computation at the edge is very minimal. Also, for this purpose, in both platforms, we generate and send the messages at an interval of 1 second. Since this workload is so lightweight, it gives us insight into the platform network overheads.
\end{itemize}

%In both Image and Audio Pipelines, we measure the end to end latency and the payload size for the messages send to cloud. In all cases, we measure the CPU and Memory usage in the raspberry pi for the duration of each benchmark. We further compartmentalize the end to end time into total compute time on the edge device, total time required to send the message though the wireless network from the edge device to the cloud, and finally the total time spend by the message in the cloud services before being written into storage.

% sep: I think that the first paragraph in Sec 3.B presents all of this information, so we don't need the table.
%\begin{table*}[]
%\begin{tabularx}{\linewidth}{|X|X|X|X|}
%\hline
%\multicolumn{1}{|l|}{}               & \multicolumn{1}{c|}{$T_1$}                                    & \multicolumn{1}{c|}{$T_2$}                             & \multicolumn{1}{c|}{$T_3$}                                   \\ \hline
%Azure Edge & UTC timestamp before sending message from IoT edge device        & UTC timestamp added to the message by IoT Hub      & UTC timestamp of creation of the Azure blob with results \\ \hline
%AWS Greengrass & UTC timestamp before sending message from Lambda Function & UTC timestamp added to the message by AWS IoT core & UTC timestamp of creation of the S3 blob for a particular message     \\ \hline
%\end{tabularx}
%\caption{Different timestamps in the edge pipelines of AWS and Azure}
%\label{table.timing_table}
%\end{table*}

\begin{figure}[]    
      \centering
      \includegraphics[width=0.9\linewidth]{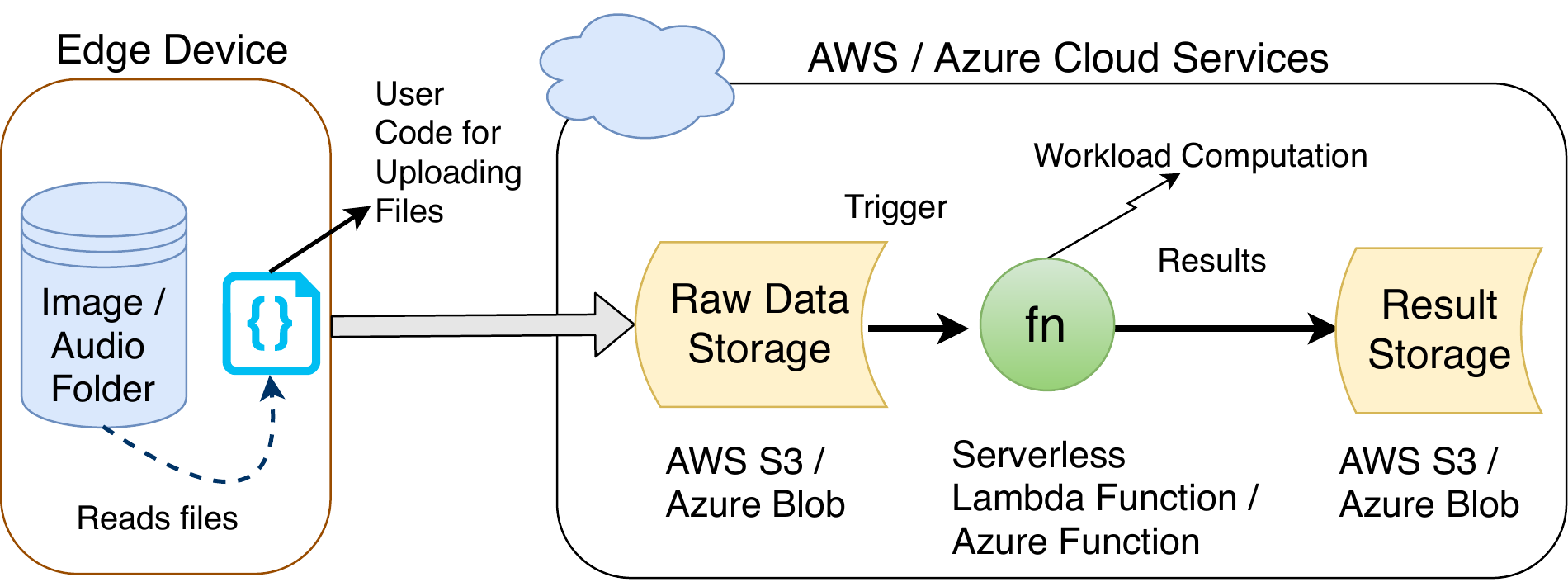}
   \caption{Schematic of a cloud only pipeline using AWS/Azure}
	\vspace{-15pt}\label{fig.Cloud_Architectures}
\end{figure}
\subsection{Metrics} 
In all pipelines of both platforms,
each message receives three UTC timestamps during the pipeline execution, as shown in 
%Table \ref{table.timing_table} and 
Fig. \ref{fig.Greengrass_architecture} and \ref{fig.azure_iot_architecture}, from which we calculate delay metrics. Here, for valid calculations on these metrics, we need the time of both the edge device and cloud to be independently synced with accurate clocks.
The user code in the edge device adds the $T_1$ timestamp before sending the result message from the edge device. 
Timestamp $T_2$ is added automatically by the platform when the message is en-queued in the IoT Hub, and finally, $T_3$ is the creation timestamp of the blob file in which the message is stored after it is routed out of the IoT Hub. 

We capture the following metrics to study performance:
\begin{itemize}[topsep=0mm,leftmargin=*]
\item \textbf{Compute time:} This is the total time required for processing one image or audio or to generate the scalar values in the Raspberry Pi and is denoted by $C_{edge}$.

\item \textbf{Time-in-flight:} This is the  time taken for a message to reach the IoT Hub  after it is sent from the edge device. It is given by $T_2 - T_1$. 

\item \textbf{End-to-end latency:} This is the difference between the time when the input is ingested at the edge device and the time when the final results are available in the storage. This value is given by $C_{edge} + (T_3 - T_1)$. 

\item \textbf{Payload size}: This is the size of the message sent from the edge device without the framework overhead.
%In case of cloud only pipelines, this is the average size of the files send to cloud over the wireless network.

\item \textbf{CPU and memory utilization}:  As the platforms use different architectures, i.e., Lambda vs Docker, it is interesting to look at the memory and CPU usage patterns. 
We measure the average CPU and memory utilization on the edge device over the execution of a given benchmark. For AWS, we use the \texttt{top} command in Linux and for Azure we use the \texttt{docker stats} command to obtain the resource utilization while the applications are running. 

\end{itemize}

Our applications log compute time, the payload size, and resource utilization locally on the edge device, while $T_1$ is added to the header. Therefore, $T_1$, $T_2$ and $T_3$ are retrieved from the message meta-data in the cloud.

\subsection{Cloud-Only Pipelines}
We implement cloud-only versions of the three benchmark workloads described in Sec.~\ref{benchmark_apps.sec} for both the Amazon and Microsoft Azure cloud platforms.
The  pipelines use the serverless architecture, as shown in Fig.~\ref{fig.Cloud_Architectures}, to process device data. All code is written in Python.

For the image classification and speech-to-text benchmarks, we upload either image or audio file from the edge device to a S3 bucket using \texttt{boto3} library.
For the scalar pipeline in Amazon AWS, we generate and upload the sensor values as JSON blob files in S3. Lambda functions are triggered by the creation of the new blob file in the bucket. The Lambda function reads the value from the blob file and simply stores it in another S3 bucket.
  The upload of a file triggers the Lambda function which, in turn, either performs the image recognition or the audio-to-text conversion.  After the computation, the results are stored as blobs in another S3 bucket. 
In the Azure implementations, we use Azure Functions, which are similar to Lambda functions.  The Azure functions are triggered by the upload of an audio or image file or a scalar value JSON file in Azure blob storage. After the 
%audio-to-text translation, the image recognition, or scalar value 
computation, the results are stored in a different blob. 
The majority of the code in cloud and edge pipelines are the same, excepting  changes  for handling input/output and receiving and handling the events due to different API specifications. 

Note that in our benchmarks, Azure functions run on Windows while Lambda functions run on Amazon Linux. Linux is available on a preview basis for Azure Functions, but it supports only JavaScript and .Net runtime as of now. Moreover, even in Windows in Azure Functions, the support of Python is in the experimental phase. 
We manually installed MXNet, OpenCV, PocketSphinx and other necessary libraries from the KUDU console in Azure Python runtime to use the libraries in Azure Functions.In AWS, the dependencies are packaged along with the Lambda function.

%%%%%%%%%%%%%%%%%%%%%%%%%%%%%%%%%%%%%%%%%%%%%%%%%%%%%%%%%%%%%%%%%%%%%%%%%%%%%%%%
\section{EXPERIMENTS} \label{sec.setup_results}
\subsection{Experimental Setup} We run the set of benchmarks using a standard Raspberry Pi 3B model as the edge device. 
%Benchmark task codes have been developed in Python. 
The edge device is connected to the internet via a wireless router using 2.4 Ghz spectrum. We have used a dedicated Stratum 1 NTP time server, TM2000A~\cite{tm2000A} with accuracy $\approx$ 50\;$\mu$s to synchronize the time of the Raspberry Pi. Also, AWS and Azure are known to use highly precise clocks to accurately sync their services.
%\das{We have not synchronized the Pi with AWS or Azure servers, Pi's time is synced with the TM2000A, for obtaining accurate time from GPS satellites.} \mw{We need to say how we synchronized the Pi to the cloud servers since we're using timestamp differences to compute delay.} 
We use AWS and Azure cloud services in the US East region, both in Virginia. We used \texttt{ping} to find the round trip latency from our institution's server to virtual machines in both Azure and AWS. We are unable to measure this from the edge device, due to security restrictions on \texttt{ping}. The average delay for AWS is 9.5\;ms and for Azure is 11.36\;ms. Assuming the extra delay within the institution network is same for both Greengrass and Azure Edge, we observe latencies to both the cloud platforms are close.
% Both correspond to data centers in Virginia, and so the latencies between our edge device and the cloud are comparable for both platforms $\approx$ 19-32\; ms.
%\mw{The above statement would be a lot stronger if we could say what the latencies to the datacenters actually are.} \das{I am not sure how to measure that.}

%The time server is set up inside the same subnet as the edge devices for low latency access. 
We use Greengrass core version 1.5.0 and Azure IoT Hub Device client 1.4.0.
In the experiments with Greengrass, each Lambda function is provisioned with 256\;MB RAM and made `long lived', i.e., it will run indefinitely. 
However, this option is absent in Azure Edge. 
In Greengrass, the image, audio and local statistics directories for storing metric values are mounted into the execution environment as `Local Resources'. In Azure  Edge, the same directories are directly mounted as volumes in the Docker container. 
In Azure Edge, we use the geo-redundant storage option RA-GRS for blob storage in the cloud. 
AWS S3 replicates data automatically across at least 3 availability zones.

We also measure performance of the cloud-only pipelines in AWS and Azure. For AWS, the memory of the Lambda functions  is set to 3008\;MB. In AWS, CPU allocation is proportional to the memory, hence, this configuration has the highest memory and CPU performance. 
As Lambda CPU is not configurable,
%Since the CPU cannot be configured in AWS Lambda, choosing specific CPU/memory tiers in Azure does not make sense. 
to keep the setups comparable, we select the Consumption Host Plan for Azure functions that auto-scales Azure functions based on system load.
In all cloud pipelines, we wait for a period of 10 to 15\;s between uploading subsequent image/audio files to avoid congesting the system.
Uploading too many image/audio files very quickly resulted in many functions being triggered concurrently and out of order.
%\sq{too late for what?}.
This results in reordering of results and some missed images, making it difficult to find the end-to-end latency.
The input data sizes are shown in Table \ref{table.bandwidthusage_total}.  
%For the image pipelines for both cloud and edge, the size is 71.69 MBytes for 500 images;  
%for the audio pipeline it is 8.83 MBytes for 104 audio files, for the scalar pipeline,  the input size is 0.05 MBytes for 200 scalar values, generated as text.

\subsection{Results}
\begin{figure}[]
      \centering
      \includegraphics[width=0.75\linewidth]{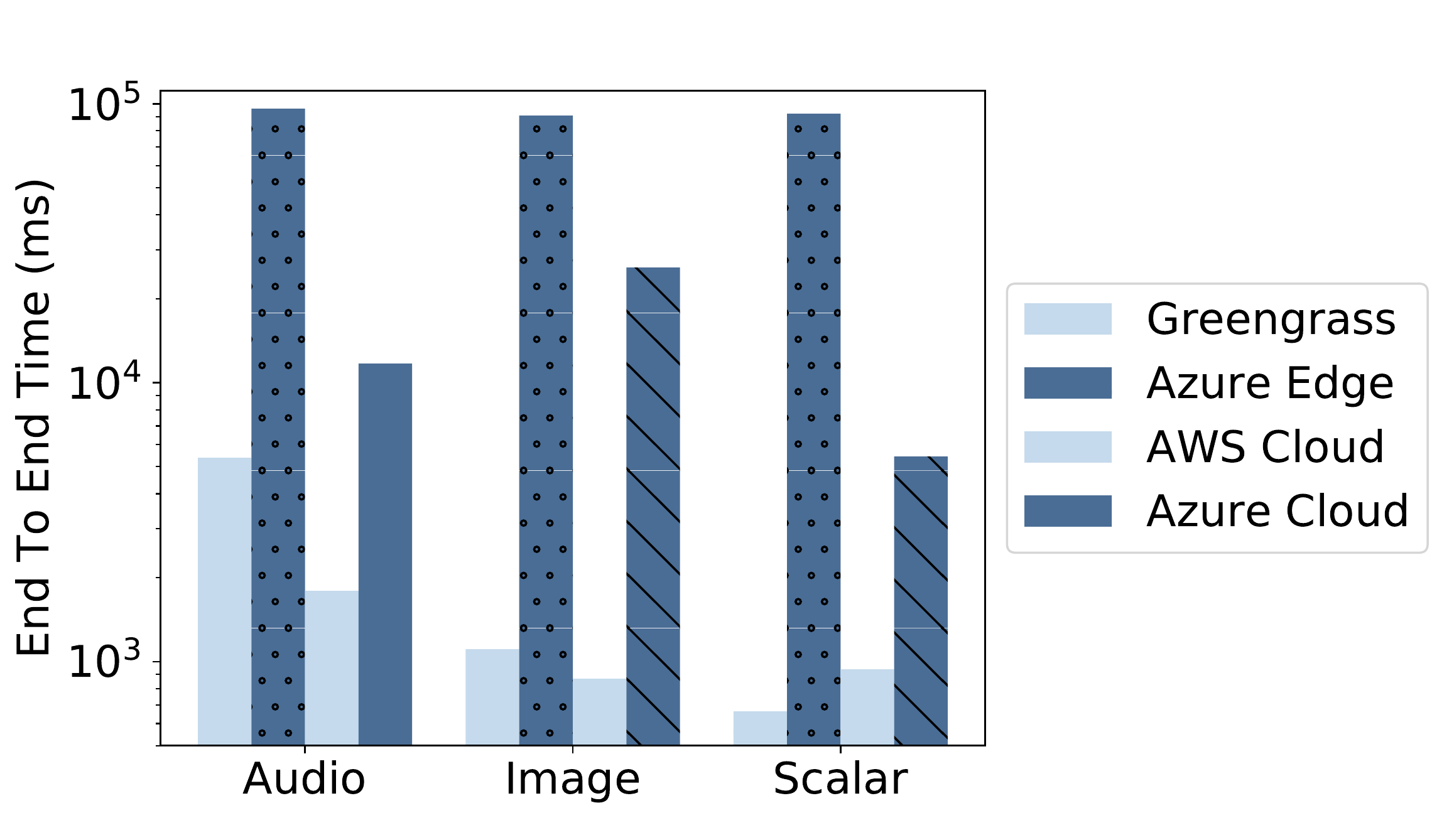}
   \caption{Average end-to-end latency in the edge and cloud-only pipelines for all benchmark applications.}%\mw{You should reorder the results to keep a consistent order of workloads: scalar, image, audio. Also you should use pattern for your bars, so that the result can be distinguished in a grayscale paper print.}
         \vspace{-10pt}\label{fig.endtoend_time}      
\end{figure}
%\begin{figure}[t]
%	\begin{subfigure}{0.5\linewidth}     
%      \centering
%      \includegraphics[width=0.95\linewidth]{Images/CPU_utilization}
%      \label{CPU_util}
%   \end{subfigure}%
%   \begin{subfigure}{0.5\linewidth}
%      \centering
%      \includegraphics[width=0.95\linewidth]{Images/Memory_utilization}
%      \label{Mem_util}
%   \end{subfigure}
%   \label{fig.local_resource}
%   \caption{Local Resource usage of edge device while using Greengrass/Azure IoT Edge services for different pipelines}
%\end{figure}

\subsubsection{End to End Latency}\label{subsec.endtoendlatency} We give the average end-to-end latency for each of the various benchmark configurations in Fig. \ref{fig.endtoend_time}. 
%End to end latency consists of compute time on device, message in flight time and time spend on cloud. 
We observe that in all three applications, across both the cloud and edge pipelines, Azure Edge has the largest end-to-end latency.
This is because  the Azure IoT service batches the messages from the edge device in the IoT Hub in the cloud, and it writes the results from the multiple messages in a batch in a single blob file. We also found that when the batching interval is 60\;s, the average time spend in hub is $\approx$ 90\;s, while for a 90\;s batching interval, the average time in the hub is $\approx $ 93-94\;s. It appears that messages are held back in the IoT hub in Azure for some time interval before being written to the blob file, and this time does not coincide with the batching interval.
If this were not the case, given that the messages are received by the  IoT Hub approximately uniformly across a batching interval, the average time a message spends in hub should have been roughly equal to half the batching interval.
This extra delay, adding to latency, exists irrespective of the blob storage type used. % whether locally redundant or geo-redundant. 

%AWS writes one blob file per message from edge device as soon as it arrives and the average time spend in the hub is around 0.55s. However, compared to AWS the batching method of Azure decreases the storage use. 

We observe that the end-to-end latency for Azure cloud pipelines is larger than both AWS edge and cloud pipelines across all applications.  
The majority of the latency is caused by the total time of execution of Azure function in the cloud.  
%\mw{how can we know that? point us to the data.}
%\sq{I think I understand what Mike is asking now. We don't have a figure or table that shows the compute time in the cloud (like Fig 5.b for the edge pipelines)I don't think we have space for another figure. Maybe you can just state the average compute time for the AWS  and Azure cloud pipelines in the text here?} 
Though, the average time for audio to speech and image recognition in Azure cloud is 5.57\;s and 1.19\;s respectively, for each message, added to this is the time for loading libraries and trigger the function, which is very high for Azure python runtime.
This may have been be caused by the fact that the Python runtime in Azure 
%runs on Windows and 
is still experimental and hence, not optimized. It may also be that importing the libraries takes a lot of time.
%We haven't used default language like C\# or Node.JS instead of Python to keep the user codes homogeneous across applications.
We obtain the smallest end-to-end latency results using the AWS Cloud (1.79\;s for Audio, 0.87\;s for Image, and 0.936\;s for Scalar), followed by Greengrass (5.36\;s for Audio, 1.1\;s for Image and 0.66\;s for Scalar). It appears that image processing at the edge with Greengrass is highly feasible, as both cloud and edge end-to-end latencies are very close. 

%Note, when using the same memory configuration of 256 MBytes in the Lambda function in cloud as compared to the Lambda function in Greengrass, we observe that the Lambda function running in the cloud is slower. This is caused by the fact that, for the Lambda functions running in the cloud, CPU is proportional to the memory configuration, whereas, Lambda function in Greengrass service is free to use all CPU of the edge device.
In Fig. \ref{fig.flight_time} we observe, Azure takes on average from 1- 18\;ms longer to deliver the messages to the cloud for the edge pipelines compared to AWS. The flight times of AWS and Azure edge are very close, which suggests that as long as the user selects data centers with similar latencies, the flight time will not contribute much to the difference of end-to-end message latencies of AWS compared to Azure.

%However, since they are almost comparable within each workload, as long as the user selects data centers with similar latencies, for time consuming computations, flight time does not add much difference between AWS and Azure's pipelines performance. 
Finally, Fig. \ref{fig.compute_time} shows that Azure, in general, has a higher compute time for all pipelines compared to Greengrass. The highest is the audio pipeline, with Azure Edge taking 6\;s, on average, and Greengrass taking 4.77\;s, on average. 
This difference may indicate a place where the different architectures (Lambda vs. Docker) may have made a difference. If this time is large, then it has a significant impact on end-to-end latency. We also observe that, for the audio pipeline in the edge, considering the average length of each clip is approximately 2.4\;s, it may not be possible to analyze the audio in real time using a Raspberry Pi without optimizing the code. However, using a more powerful edge device would help to reduce compute time. 

\begin{table}[]
\resizebox{0.48\textwidth}{!}{%
\begin{tabular}{|c|c|c|c|c|c|}
\hline
\multicolumn{2}{|l|}{\multirow{2}{*}{}}                                         & \multirow{2}{*}{\textbf{\begin{tabular}[c]{@{}c@{}}Total \\ Input Size\\ (Mbytes)\end{tabular}}} & \multirow{2}{*}{\textbf{\begin{tabular}[c]{@{}c@{}}Total raw\\ Payload Size\\ (Mbytes)\end{tabular}}} & \multicolumn{2}{c|}{\textbf{\begin{tabular}[c]{@{}c@{}}Total MBytes\\ Transmitted \\ in Network\end{tabular}}} \\ \cline{5-6}
\multicolumn{2}{|l|}{}                                                          &                                                                                                  &                                                                                                       & AWS                                                  & Azure                                               \\ \hline
\multirow{2}{*}{\begin{tabular}[c]{@{}c@{}}\textbf{Audio}\\ Trials = 104\end{tabular}}  & Edge  & \multirow{2}{*}{8.83}                                                                            & 0.02                                                                                                  & 0.25                                                 & 0.26                                                \\ \cline{2-2} \cline{4-6}
                                                                        & Cloud &                                                                                                  & 8.83                                                                                                  & 9.06                                                 & 9.09                                                \\ \hline
\multirow{2}{*}{\begin{tabular}[c]{@{}c@{}}\textbf{Image}\\ Trials = 500\end{tabular}}  & Edge  & \multirow{2}{*}{71.69}                                                                           & 0.38                                                                                                  & 0.9                                                  & 0.96                                                \\ \cline{2-2} \cline{4-6}
                                                                        & Cloud &                                                                                                  & 71.69                                                                                                 & 73.10                                                & 73.49                                               \\ \hline
\multirow{2}{*}{\begin{tabular}[c]{@{}c@{}}\textbf{Scalar}\\ Trials = 200\end{tabular}} & Edge  & \multirow{2}{*}{0.05}                                                                            & \multirow{2}{*}{0.05}                                                                                 & 0.33                                                 & 0.26                                                \\ \cline{2-2} \cline{5-6}
                                                                        & Cloud &                                                                                                  &                                                                                                       & 0.47                                                 & 0.38                                                \\ \hline
\end{tabular}%
}
\caption{Total input, payload, and actual data transmitted in edge and cloud only pipelines for the three benchmark applications along with number of trials.}%
\vspace{-10pt}\label{table.bandwidthusage_total}%
\end{table}%

%\begin{table}[]
%\centering
%\begin{tabular}{|l|c|c|c|}
%\hline
%       & \textbf{Audio-to-Text}  & \textbf{Image Recognition}   & \textbf{Scalar Sensor} \\ \hline
%Edge   & 0.162 KB                & 0.751 KB                     & 0.234 KB      \\ \hline
%Cloud  & 84.85 KB                & 143.12 KB                    & 0.24 KB             \\ \hline
%\end{tabular}%
%\caption{Average bandwidth usage, i.e., data sent from the edge device to the cloud, in edge vs. cloud-only pipelines.}
%\label{table.bandwidthusage_perfile}
%\end{table}

\subsubsection{Bandwidth Utilization} 
\begin{figure*}[t]
	\begin{subfigure}[t]{.25\textwidth}    
      \centering
      \includegraphics[width=\linewidth]{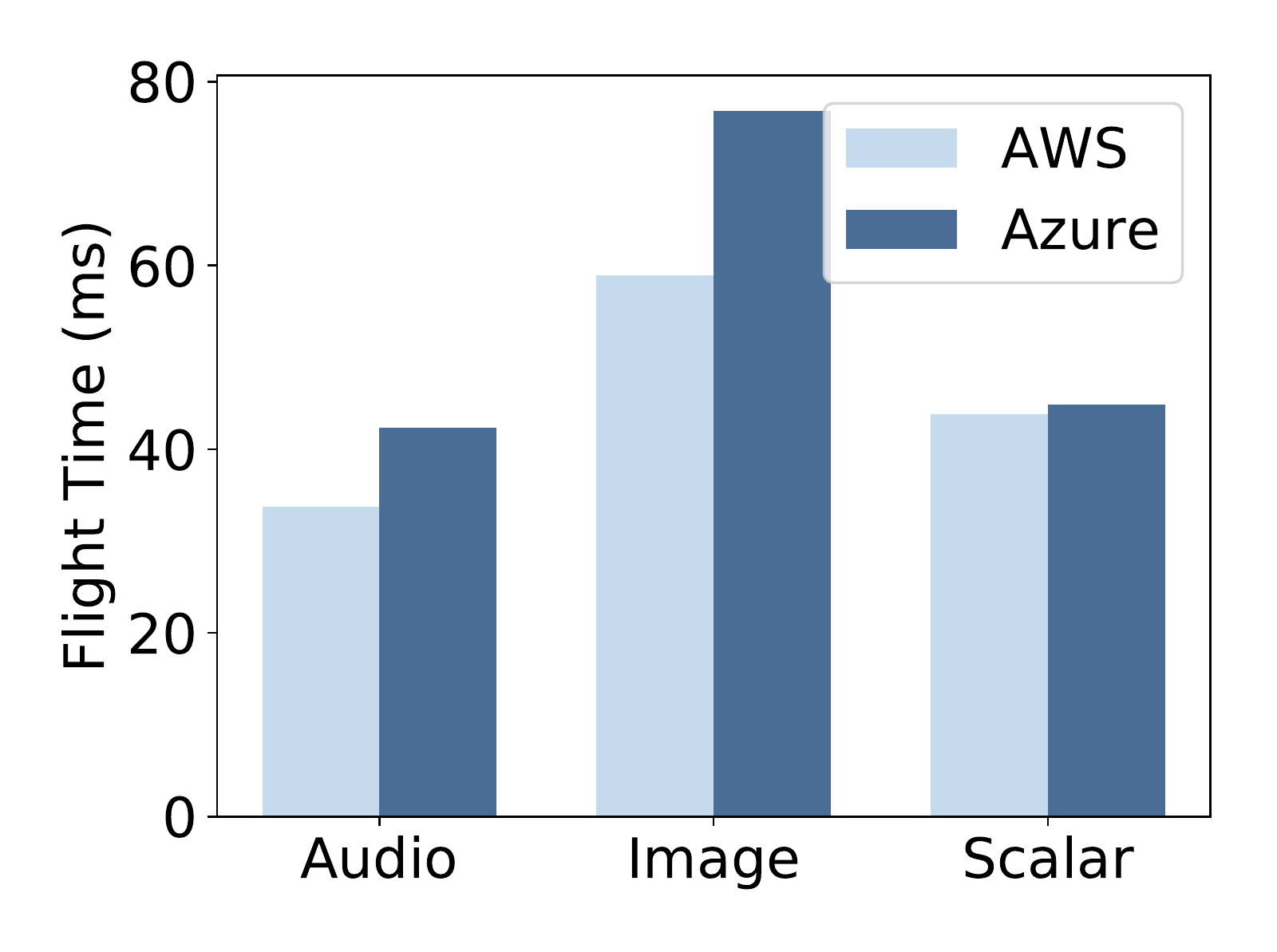}
      \caption{Avg. time-in-flight/ message.}
      \label{fig.flight_time}
   \end{subfigure}\hfill
   \begin{subfigure}[t]{.25\textwidth} 
      \centering
      \includegraphics[width=\linewidth]{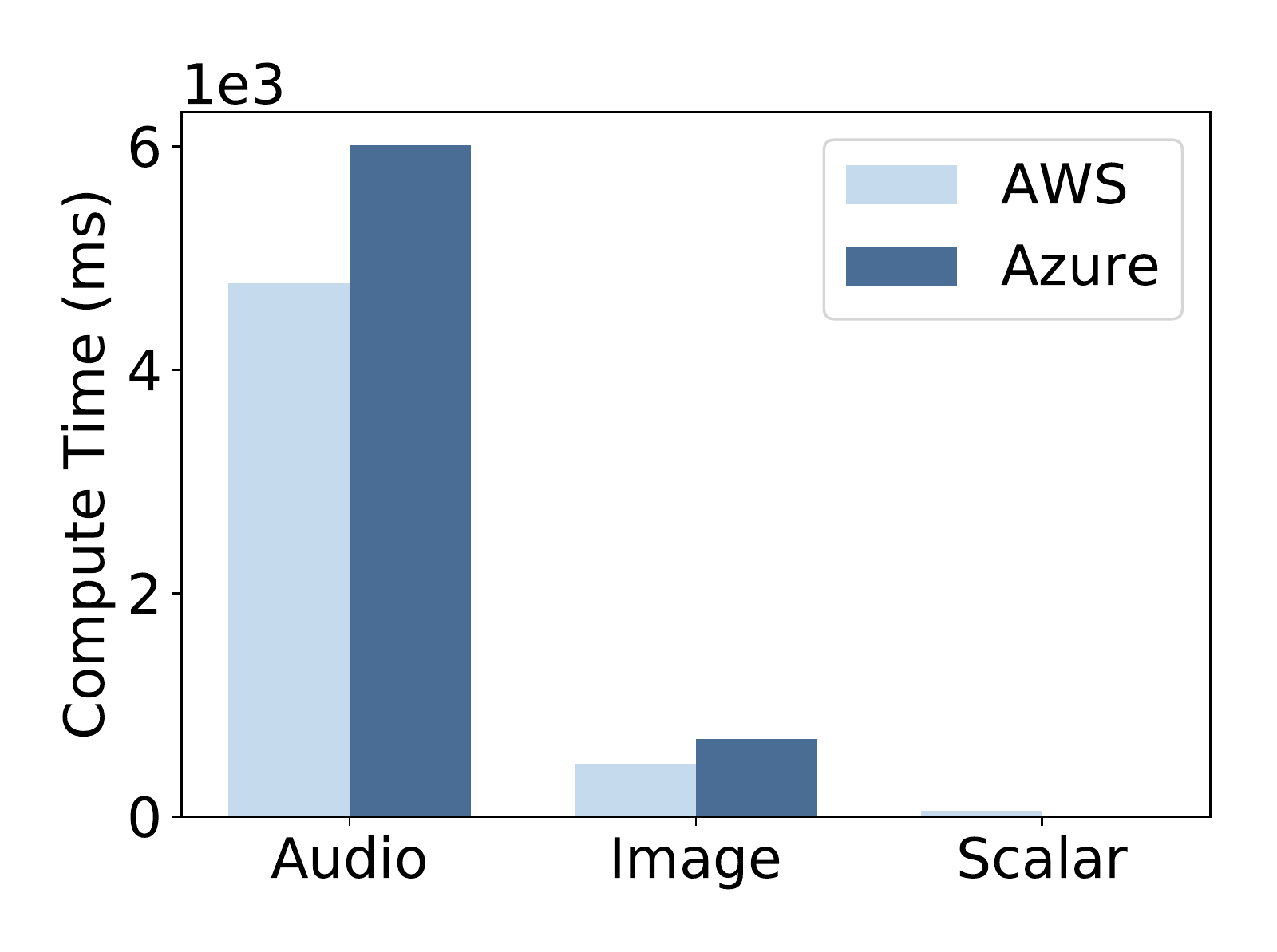}
      \caption{Avg. compute time/ message.}      
      \label{fig.compute_time}      
   \end{subfigure}\hfill
   \begin{subfigure}[t]{.25\textwidth} 
      \centering
      \includegraphics[width=\linewidth]{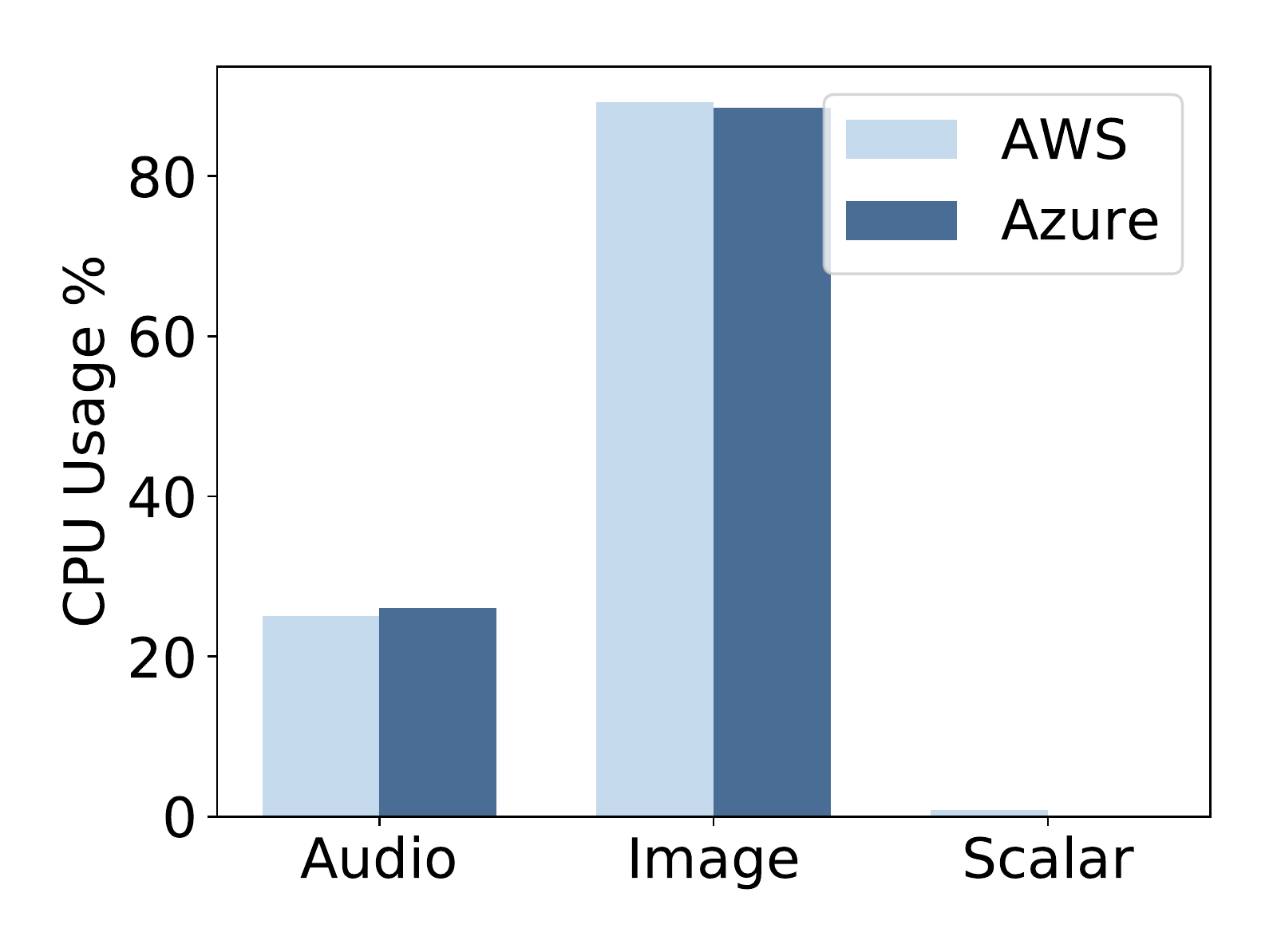}
      \caption{Avg. CPU utilization \%.}      
      \label{fig.CPU_util}
   \end{subfigure}\hfill
   \begin{subfigure}[t]{.25\textwidth} 
      \centering
      \includegraphics[width=\linewidth]{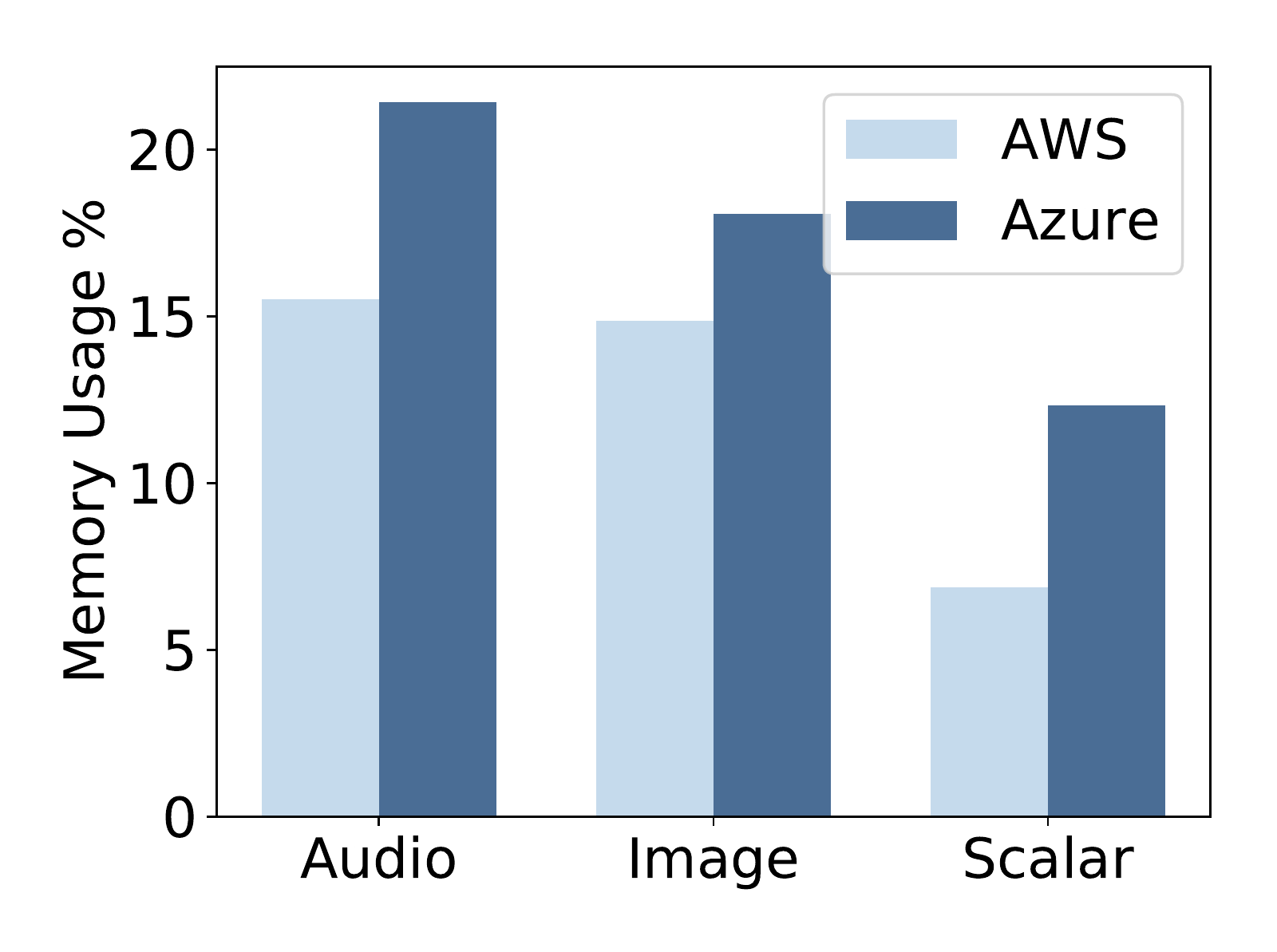}
      \caption{Avg. RAM utilization \%.}      
      \label{fig.Mem_util}
   \end{subfigure}\hfill
   \caption{Comparison of the performance metrics in Greengrass (AWS) and Azure Edge (Azure) across different pipelines.}
   \vspace{-10pt}\label{fig.aws_azure_timings}
\end{figure*}

%\note{Insert measuring network overhead using \texttt{vnstat} to measure total network usage including framework overhead? How does the framework overhead of the two frameworks differ?} 
The average payload size in both platforms for audio pipeline is 162\;bytes, for image pipeline it is 752\;bytes and for scalar it is 234\;bytes.
%This is important to study because the one of the fundamental goals of edge computing is bandwidth reduction. 
On comparing flight times for each edge pipeline in Fig. \ref{fig.flight_time} we observe that the transmission delay (flight time) of messages between the edge and cloud is roughly proportional to the message size.
A cloud-only approach requires the upload of the raw image or audio file to the cloud service. 
%Naturally this uses up a lot of bandwidth. 
In the edge pipelines, we only the send results of the applications as text to the cloud. Hence, we observe that there is drastic reduction in the per message size in edge pipelines compared to cloud. We also used \texttt{vnstat}~\cite{vnstat} %\footnote{https://github.com/vergoh/vnstat} 
to obtain the total bandwidth usage of the applications in the edge and cloud pipelines, shown in Table \ref{table.bandwidthusage_total}. The results are measured with respect to 200 scalar values, 500 images, and 104 audio files, respectively. To avoid measuring the TLS handshakes and other module startup network overhead, we explicitly add a configurable delay (60\;s, in this case) before the application begins processing data. We see that framework data overhead itself is negligible and comparable in both platforms.
Comparing the total data transmitted during pipeline executions, we see a massive reduction of data transmission while using the edge pipelines compared to cloud. AWS sends 36 times and 81 times more data when using the cloud pipelines compared to the edge, for the audio and image applications, respectively. Azure sends 36 times and 77 times more data using the cloud pipelines compared to the edge in audio and image applications, respectively.

\subsubsection{Local Resource Utilization} We study the average resource usage of the edge device (Raspberry Pi) across the edge pipelines over three separate runs of the experiments. In Azure Edge, we look at the average total CPU and memory percentage used by  all of the containers running specific to Azure Edge. In Greengrass, we look at the total average CPU and memory percentage usage by all processes under the greengrass user, \texttt{ggc\_user}. 

%All Greengrass processes including auth services,  connection manager etc runs under this user, hence we take cumulative resource utilization of that user.
We observe in Fig. \ref{fig.CPU_util}, \ref{fig.Mem_util} that the image recognition application is predominantly a CPU intensive job, with CPU utilization as high as 88-90\% in both Azure and AWS. We also observe that audio-to-text is not very CPU-heavy, but it consumes more memory than the other applications. We further observe that within each application, the CPU \% of Greengrass and Azure Edge are very similar, though the RAM consumed in Azure Edge is always higher. Azure Edge consumes on average about 29.5\;MB to 54.5\;MB more memory on the Raspberry Pi. 
We believe this difference would be less discernible if a more powerful edge device were used. Overall, we observe that the edge pipelines, on average, consume less than 200\;MB RAM and do not saturate the CPU usage. This strengthens the case for the feasibility of running some carefully chosen computations on resource-constrained devices.

%\textbf{N.B.} We don't explicitly set the ram or cpu constraints on the docker containers in Azure IoT hub and can possibly use everything available in the pi. However, in Greengrass, we explicitly set 256 ram on the Lambda functions. Yet the performance is comparable. \note{What does this say?}

\subsubsection{Infrastructure Cost} 
%To get an idea about feasibility of edge pipelines compared to cloud only pipelines,
We do a rough infrastructure cost estimate of running the applications in the edge pipelines versus the cloud-only pipelines. Cost of running pipelines in both vendors are comparable and so, for simplicity, here, we  look only at the image pipeline in AWS. 
Assume there is one traffic camera, generating one image every 10\;s. Let us further assume the  average image size  is similar those used in our image benchmarks, i.e., 143.12 KB (In general, image size and rate would be larger in real world scenarios.) 
This amounts to $6\times60\times24\times30 = 259,200$ images per month. Also, on average, the duration for which AWS charges for Lambda function execution is 300ms in the image recognition cloud only pipeline in our study for image of that size. 
 In the Greengrass image pipeline, the average size of a message to the cloud is 752\;bytes. We assume, with headers, it would be approximately 1\;KB per message. All prices are calculated in region US-East, Virginia. 

For Greengrass total cost is the expense of running Greengrass plus the cost of storing results in S3 plus put requests for results in S3, which equals $0.2627 + 0.0057 + 1.29=\;\$\;1.5584$\;/\;month. 
For the cloud pipeline, the cost is the expense of storing raw images and final results in S3 plus get and 2$\times$put requests cost in S3 plus the cost of running Lambda functions , which equals $0.814 + 0.0057 + 2.69 + 4.517 = \;\$\;8.027$ \;/\;month.  Though this estimate appears cheap, if there are, for example, 50 road side cameras, the cost for the cloud-only pipeline escalates quickly. This rough cost estimate indicates that executing image recognition on the cloud setting is $\approx$ 5.2\;x more expensive than edge, with only an extra 230 ms in average end-to-end latency. It is possible to reduce the cost of the cloud-only pipeline by using less powerful Lambda functions, however, the storage cost alone is larger than the entire edge pipeline cost. 
Bandwidth usage-wise, the edge pipeline sends around 253.125\;MB data over the network per month, whereas uploading images to cloud requires sending 35.38\;GB data/month.

\subsection{Discussion} \label{sec.discussion}
%Here, we discuss several observations and lessons learned during the benchmarking process. 
We observed that both platforms do not handle very high throughput messaging well yet. In this case either the messages are delayed or fail to reach the cloud from the edge device. In future we want to benchmark this throughput.
%This may have caused the bottle neck in the system because it is simply unable to route such a huge number of messages being published every second. The bottleneck, as and Intel Xeon E5-2630 v4 processor. 
%
%In Azure Edge, on a fresh system restart on edge device, in general, the user module starts earlier than \texttt{edgeHub} module. Hence, if user module starts sending messages before \texttt{edgeHub} fully initializes, the authentication and connection protocol setup would still be going on, and messages will be delayed or timed out. We contacted the Azure team, they are implementing a fix.

Another important consideration is the relative ease of deploying dependencies and libraries. We can package all necessary libraries in the Docker container with a single Dockerfile in Azure Edge. On the contrary, in Greengrass Lambda functions, we need to compile external dependencies in required environment and add them in a zip file for deploying. We feel that the former is a cleaner choice for adding and managing a lot of external libraries.

%\sq{Is it tricky? or just not possible. If it is tricky, give a short explanation as to why.} 

%One alternative is to manually install the necessary libraries in the edge device itself. \sq{Is there a drawback to this alternative. What should the reader take away from this paragraph? Is one framework better with libraries than the other?}

%In order to access local resources or volumes, in Greengrass, to mount a local directory in the Lambda function execution environment, user has to manually create the mount directory in the edge device as well. So, both the actual and mount directory needs to be present. This need user intervention every time the user want to change the mount directory based on code requirement. This is easier in docker where the mount path is auto created inside the docker container and can be configured during deployment. 

In the end, from our experience, we feel both these approaches are suitable for carrying out edge computation. However, the higher end-to-end latency of Azure may be a problem for latency sensitive applications.
%Which platform to choose, ultimately boils down to the developer preference and comfort. 
Although Azure has richer customization options, we found development and integration to be easier in the AWS Greengrass platform.
%\sq{I think we should not overlook the higher end-to-end latency in Azure Edge because of batching. Can you add a sentence about the pros and cons of batching (are there any pros?)}

%%%%%%%%%%%%%%%%%%%%%%%%%%%%%%%%%%%%%%%%%%%%%%%%%%%%%%%%%%%%%%%%%%%%%%%%%%%%%%%%
\section{CONCLUSION} \label{sec.conclusion}
We have presented EdgeBench, a benchmark suite for serverless edge computing plaforms.
With this suite, 
we have studied two managed edge computing platforms, Greengrass and Azure Edge. Further, we have compared the performance of these platforms with that of cloud-only implementations of the same benchmarks. Our results show that the performance of Greengrass and Azure Edge are comparable, with the exception
that Azure Edge exhibits higher end-to-end latency due to its batch-based processing approach. Further, our results show that for the image and scalar pipelines, the performance of Greengrass is comparable to that of the AWS cloud-only pipelines, while reducing the network bandwidth usage. These results indicate that edge computing is a promising alternative to cloud computing for CPU light workloads.  In future work, we plan to extend EdgeBench to include additional applications and edge platforms. 
%%%%%%%%%%%%%%%%%%%%%%%%%%%%%%%%%%%%%%%%%%%%%%%%%%%%%%%%%%%%%%%%%%%%%%%%%%%%%%%%
%\section{ACKNOWLEDGMENTS}
%

\addtolength{\textheight}{0cm}   % This command serves to balance the column lengths
                                  
%%%%%%%%%%%%%%%%%%%%%%%%%%%%%%%%%%%%%%%%%%%%%%%%%%%%%%%%%%%%%%%%%%%%%%%%%%%%%%%%

\bibliographystyle{ieeetr}
\bibliography{biblio} 
\end{document}